\documentclass[10pt,a4paper,newlfont,twocolumn,aps,pra,superscriptaddress]{revtex4}
\def\hs{\hspace{0.5cm}}
\def\ea{{\it et al.}}
\def\ft{\hspace{2mm}}
\def\xhe4{$^4$He}
\def\aho{$a_{ho}\,=\,\sqrt{\hbar/(m\omega_{ho})}$}
\def\bls{\footnotesize\baselineskip 0.1cm}
\usepackage[sort&compress]{natbib}
\usepackage[pdftex]{graphicx} 
\pdfoutput=1
\begin{document}

\title{Self-interfering matter-wave patterns generated by a moving laser obstacle 
in a two-dimensional Bose-Einstein condensate inside a power trap cut off 
by box potential boundaries}
\author{Roger R. Sakhel}
\affiliation{Department of Basic Sciences, Faculty of Information Technology, 
Isra University, Amman 11622, Jordan}
\author{Asaad R. Sakhel}
\affiliation{Department of Applied Sciences, Faculty of Engineering 
Technology, Al-Balqa Applied University, Amman 11134, Jordan}
\author{Humam B. Ghassib}
\affiliation{Department of Physics, The University of Jordan, Amman, Jordan}
\begin{abstract}
We report the observation of highly energetic self-interfering matter-wave
(SIMW) patterns generated by a moving obstacle in a two-dimensional 
Bose-Einstein condensate (BEC) inside a power trap cut off by hard-wall 
box potential boundaries. The obstacle initially excites circular dispersive 
waves radiating away from the center of the trap which are reflected from 
hard-wall box boundaries at the edges of the trap. The resulting interference 
between outgoing waves from the center of the trap and reflected waves from 
the box boundaries institutes, to the best of our knowledge, unprecedented 
standing wave patterns. For this purpose we simulated the time dependent 
Gross-Pitaevskii equation using the split-step Crank-Nicolson method. 
The obstacle is modelled by a moving impenetrable Gaussian potential barrier. 
Various trapping geometries are considered. 
\end{abstract}
\date{\today}
\maketitle

\section{Introduction}

\hs Recently, there has been considerable interest in the investigation
of the effects of a moving obstacle (or object) through a Bose-Einstein
condensed system in various trapping geometries 
\cite{Jackson:1998,Engels:2007,Onofrio:2000,Gladush:2009,Winiecki:1999,
Horng:2009,Susanto:2007,Carusotto:2006,Astrakharchik:2004}. 
The obstacle is a potential barrier generated by a Gaussian laser beam 
\cite{Onofrio:2000,Carusotto:2006}, which can be repulsive or attractive 
depending on the wavelength of the beam. Previous work has also shown 
that the motion of this obstacle, whether linear or rotational, causes 
excitations leading to strikingly interesting phenomena such as vortices 
\cite{Jackson:1998,Caradoc:1999,Caradoc:2000,Gonzalez:2007,Madison:2000,Raman:2001}, 
solitons \cite{Ma:2010,Gonzalez:2007}, crescent vortex solitons \cite{He:2008}, 
and dispersive waves accompanying the solitons \cite{Hakim:1997}. In addition,
there has been growing interest in the investigation of $\check{\hbox{C}}$erenkov 
radiation \cite{Susanto:2007,Carusotto:2006} and waves generated by an obstacle
moving with supersonic velocity \cite{Horng:2009,Gladush:2009}. 
However, to the best of our knowledge, none of the previous 
literature considered the effects of adding a hard-wall to the magnetic 
trap on the dynamics of the BEC excited by a moving obstacle.
It is well known, that vortices can be excited by moving obstacles
inside a trapped BEC. However, the parameters used in the present
investigation are lower than what is required to obtain significant
vortex features, such as the vortex-antivortex pairs observed by
Jackson \ea\ \cite{Jackson:1998}.

\hs One goal of this work is to check for a possible existence of 
vortices in our systems below. Our motivation for this stems from 
the vast literature on vortices in BECs confined by various traps, 
with interesting and exciting phenomena both in theoretical 
\cite{Caradoc:1999,Feder:1999,Caradoc:2000,Adhikari:2002,Vignolo:2007,Goldbaum:2009,Salasnich:2009,Davis:2009,Adhikari:2010}
and experimental domains \cite{Inouye:2001,Tung:2006,Williams:2010,Neely:2010}.

\hs One way to explore trapped BECs excited by a moving obstacle
is by the time-dependent Gross-Pitaevskii equation (TDGPE) whose
solutions are known to be solitons \cite{Nonnenmacher:1983} 
depending on the interactions and certain boundary conditions; the
nonlinearity in the GPE being the cause for the appearance of these 
excitations. The role of the nonlinearity in the generation of 
solitonic solutions for the nonlinear Schr\"odinger equation has 
been studied as early as 1983, when Nonnenmacher and Nonnenmacher 
\cite{Nonnenmacher:1983} found, that for repulsive interactions no 
solitons are observed; unlike the case for attractive interactions.

\hs In this work, the split-step Crank-Nicolson method 
\cite{Muruganandam:2009} is applied to numerically solve the 
two-dimensional (2D$^{\hbox{al}}$) TDGPE, involving a harmonic or 
(power-law) PL trap plus a moving obstacle. The trap is cut off 
by hard-wall box-potential (HWBP) boundaries serving as reflectors 
of matter waves. In fact, the box potential, along with other hard-wall
trapping geometries, has been used previously by Ruostekoski \ea\
\cite{Ruostekoski:2001}, who explored the interference of a 
BEC in a hard-wall trap plus the nonlinear 
Talbot effect. We shall comment on their work in the discussion 
section. According to Ruostekoski \ea, the hard-walls can be realized 
experimentally by a blue-detuned light sheet. Consequently, the trap
is limited by the hard walls and an expanding BEC can be reflected 
off these walls producing a self-interfering matter-wave (SIMW) 
field. Essentially, then, we propose an experiment similar
to that of Ruostekoski \ea, in which we use a 2D$^{\hbox{al}}$ square box potential.
But instead of switching off the isotropic magnetic trap in order
to allow the BEC to expand, we excite the BEC by a moving obstacle
in order to generate dispersive waves (DWs) that travel towards
the hard wall in order to get reflected. Thus, if no DWs are emitted,
no SIMWs can be obtained. However, our 2D$^{\hbox{al}}$ square box is achieved 
numerically by a different method than that of the latter authors. 
The main motivation of this work, then, is largely based on the latter work
of Ruostekoski \ea\ \cite{Ruostekoski:2001}. Additional motivation
stems from a previous investigation by Jackson \ea\ \cite{Jackson:1998}, 
who simulated the motion of an ``object" through a dilute BEC in 
a 2D$^{\hbox{al}}$ harmonic trap, where the object (or obstacle) 
corresponds to a laser-induced vortex potential. Our goal is to 
explore the same system of Jackson \ea, at a lower mean-field interaction 
strength ${\cal N}$ but with an additional box-potential trap. We 
essentially take a closer look at the dynamics of the density 
surrounding the central BEC at distances larger than $\sim 4$ trap 
lengths from the center of the trap in the event of matter-wave
reflections. Another motivation comes from the work of 
Hakim \cite{Hakim:1997}, who investigated the superflow of a 
nonlinear Schr\"odinger fluid past an obstacle. He found that the 
obstacle repeatedly emits gray solitons which propagate downstream; 
whereas DWs propagating upstream are emitted at the same time. 
Since the calculations of Ref.\cite{Hakim:1997} were conducted 
in 1D and outside a trap, the importance of our current investigation 
lies, then, in exploring these systems in 2D and further 
inside various trapping geometries inside a hard wall potential. 
It is, in fact, the DWs which cause the phenomena obtained in our 
current study. To this end, we chiefly aim at presenting a 
phenomenological investigation of SIMW patterns generated by a moving 
obstacle inside a 2D$^{\hbox{al}}$ BEC in various trapping geometries 
surrounded by a HWBP.

\hs Several questions are tackled: First, can we observe DWs in 
2D$^{\hbox{al}}$ harmonically or nonharmonically-trapped BECs 
excited by a moving obstacle, such as was observed by Hakim 
\cite{Hakim:1997} for the case of a 1D uniform Bose gas? Second, by
adding the HWBP to the harmonic or PL trap, what new features are 
observed when the latter matter waves are reflected from the hard 
walls? Will they be similar to Ref.\cite{Ruostekoski:2001}? Third, 
how does the energy of the system behave with time? Fourth, what 
about the momentum density dynamics?

\hs Further, additional simulations of the 1D$^{\hbox{al}}$ GPE are 
conducted for the same systems above in an attempt to reproduce the 
dispersive waves observed by Hakim \cite{Hakim:1997}. The purpose is 
to compare these 1D$^{\hbox{al}}$ disturbances with the 2D$^{\hbox{al}}$ 
disturbances in order to understand their nature in 2D. 

\hs Our key results are as follows. i) It is found that in a 
2D$^{\hbox{al}}$ harmonic or PL trap, highly-energetic DWs can 
be emitted from the moving obstacle. These DWs are highly energetic, 
capable of surmounting the PL potential barrier to far distances from
the trap center. Upon reflection of their wave fronts from the HWBP, 
their interference with incoming DWs produces at some point
perpendicular, 2D$^{\hbox{al}}$ density-stripes in the matter-wave field, 
largely influenced by the artefacts of the trapping geometry. This phenomenon 
has not, to the best of our knowledge, been reported elsewhere to 
this date; ii) the accompanying energy dynamics of these systems 
display an oscillatory pattern indicative of a possible presence 
of solitons; iii) when the trapping strength of the PL is increased, 
the energy otherwise expended on the creation of the SIMWs is 
channeled to excite a larger number of solitons inside the trap, 
eventually leading to a ``soliton gas"; iv) the condensate dynamics 
reveal further an oscillatory behaviour signalling that particles 
are excited out of and deexcited back into the BEC.

\hs The paper is organized as follows. In Sec.\ref{sec:method}, we 
briefly describe the method used. In Sec.\ref{sec:results}, we 
present our results, and in Sec.\ref{sec:discussion} we discuss
them. The conclusion is presented in Sec.\ref{sec:conclusion}

\section{Method}\label{sec:method}

\subsection{Time-dependent Gross-Pitaevskii Equation}

\hs In this work, we apply the split-step Crank-Nicolson method 
\cite{Muruganandam:2009} to numerically solve the 2D$^{\hbox{al}}$ 
TDGPE given by

\begin{eqnarray}
i\hbar\frac{\partial \Psi(\mathbf{r};\tau)}{\partial \tau}\,&=&\,
\left[-\frac{\hbar^2}{2 m}\left(\frac{\partial^2}{\partial x^2}\,+\,
\frac{\partial^2}{\partial y^2}\right)\,+\,V(x, y; \tau)\,+\,\right.\nonumber\\
&&\left.g N |\Psi(\mathbf{r};\tau)|^2\right]\Psi(\mathbf{r};\tau),
\label{eq:2D-GPE}
\end{eqnarray}

where $N$ is the number of particles, $g\,\equiv\,4\pi\hbar^2 a_s/m$ is the
interaction parameter, with $a_s$ the s-wave scattering length in the
low-energy and long-wavelength approximation, $m$ is the mass of the
boson, and $\hbar$ is Planck's constant. For the present purposes, 
$V(x, y; \tau)$ is taken as a potential which varies with time. 
According to Muruganandam and Adhikari (MA) \cite{Muruganandam:2009}, 
the TDGPE can then be recast into a dimensionless form using units 
of the trap such that

\begin{eqnarray}
&&\left[-\frac{1}{2}\frac{\partial^2}{\partial x^2} 
-\frac{1}{2}\frac{\partial^2}{\partial y^2} +
\tilde{V}(x, y; t) + \right.\nonumber\\ 
&&\left.{\cal{N}}|\varphi(x,y;t)|^2 -
i\frac{\partial}{\partial t} \right]\varphi(x,y;t)\,=\,0, 
\label{eq:rescaled-2DTDGPE}
\end{eqnarray}

where $t\,=\,\omega_{ho}\tau$ is a unitless time, 
$\tilde{V}(x, y; t)\,=\,V(x, y; t)/(\hbar\omega_{ho})$ and 
$x\rightarrow x/\ell$, $y\rightarrow y/\ell$ with $\ell\,=\,\sqrt{\hbar/(m\omega_{ho})}$
the trap length. The parameter ${\cal N}$ is

\begin{equation}
{\cal{N}}\,=\,\frac{2 N a_s}{\ell}\,\sqrt{2 \pi \lambda},
\label{eq:rescaled-interaction-parameter-N}
\end{equation}

$\lambda$ being a parameter describing the width of the ground-state 
wave function $\phi_0(z)\,=\,(\lambda/\pi)^{1/4} \exp(-\lambda z^2/2)$
in the $z$-direction, which was then integrated out from the 3D$^{\hbox{al}}$ 
wave function so as to get the 2D$^{\hbox{al}}$ form 
Eq.(\ref{eq:rescaled-2DTDGPE}) above. That is, we consider the $z$-direction 
to have harmonic confinement of infinite strength. Note further that MA 
rescaled the wave function using 
$\varphi(x, y; t)\,=\,\sqrt{\ell^3}\Psi(\mathbf{r};\tau)$.

\subsection{Trapping potential and moving obstacle}\label{sec:trapping-potential-and-moving-obstacle}

\hs The time-dependent trapping potential is composed of two parts: 
a static external general 2D$^{\hbox{al}}$ PL trap $\tilde{V}_{PL}(x, y)$, 
and a moving potential (MP) barrier (or obstacle) with time  
$\tilde{V}_{MP}(x, y; \tau)$ which is generated 
by sweeping a blue-detuned laser beam through the BEC (see, e.g., 
\cite{Matthews:1999}). Hence, 
$\tilde{V}(x, y; t)\,=\,\tilde{V}_{PL}(x, y)\,+\,\tilde{V}_{MP}(x, y; t)$.
The PL potential $\tilde{V}_{PL}(x, y)$ is taken to be of the form

\begin{equation}
\tilde{V}_{PL}(x, y)\,=\,\sigma (|x|^{p_1}\,+\,\kappa |y|^{p_2}),
\label{eq:PL-potential}
\end{equation}

where $p_1,\,p_2\,>0$ can be any positive numbers, $\kappa$ is an 
anisotropy parameter, and $\sigma$ is the strength of the PL potential,
taken here to be 1. Thus for $p_1=p_2=2$ we have the usual harmonic 
trap, whether isotropic ($\kappa=1$) or anisotropic ($\kappa > 1$ or $< 1$). 
Both $\sigma$ and $\kappa$ are unitless, since $x$ and $y$ are unitless
as well. The goal of using Eq.(\ref{eq:PL-potential}) is to
explore the effect of different curvatures of the external trapping 
potential on the dynamics of a trapped BEC when excited by a moving
obstacle.

\hs Next, in order to excite the BEC, we follow Jackson \ea\ 
\cite{Jackson:1998} and Bao and Du \cite{Bao:2004} and introduce the
obstacle in the form of a velocity-dependent Gaussian potential given by

\begin{equation}
\tilde{V}_{MP}(x, y; t)\,=\,A \exp[-\beta x^2\,-\,\beta (y\,-\,vt)^2],
\label{eq:velocity-dependent-Gaussian-potential}
\end{equation}

$A$ being the amplitude, $v$ the velocity in the $y-$direction, and 
$\beta$ the exponent determining the width. 

\hs We further cut off the harmonic or PL trap by a HWBP, which can
be numerically achieved by forcing the wave function of the system
to vanish at the hard walls. This can be realized by imposing the
following boundary conditions

\begin{eqnarray}
\varphi(x,-L_y/2;t)\,=\,\varphi(x,L_y/2;t)\,=\,0\nonumber\\
\varphi(-L_x/2,y;t)\,=\,\varphi(L_x/2,y;t)\,=\,0.\nonumber\\
\label{eq:boundary-conditions}
\end{eqnarray}

Thus the overall trapping geometry is then described by

\begin{eqnarray}
&&\tilde{V}_T(x,y;t)\,=\,\nonumber\\
&&\left\{\begin{array}{l@{\quad :\quad}r}
\tilde{V}(x,y;t) & \displaystyle -\frac{L_x}{2}< x<\frac{L_x}{2}; -\frac{L_y}{2}< y<\frac{L_y}{2} \\
\infty & x\,=\,\pm \displaystyle \frac{L_x}{2}, y\,=\,\pm\frac{L_y}{2}.
\end{array}\right.\nonumber\\
\label{eq:total-trapping-geometry}
\end{eqnarray}

We take $L_x\,=\,L_y\,=\,20$ in trap units. Note that this is
different from the way the box potential was constructed in 
Ref.\cite{Ruostekoski:2001}, where a set of large-amplitude
Gaussians was used. In essence, we mimic here an ``exact" 
hard-wall potential where the BEC cannot tunnel $-$even slightly$-$
through the wall.

\subsection{Energy functional}

\hs The energy is evaluated by the well-known energy functional in 
trap units

\begin{eqnarray}
E(t)\,&=&\,\int d^2\mathbf{r} \left[\frac{1}{2}|\nabla\varphi(x,y;t)|^2\,+\,
\tilde{V}(x,y;t)|\varphi(x,y;t)|^2\,+\,\right.\nonumber\\
&&\left. {\cal{N}} |\varphi(x,y;t)|^4\right].
\label{eq:energy-functional}
\end{eqnarray}

Since the above functional includes the moving obstacle 
Eq.(\ref{eq:velocity-dependent-Gaussian-potential}), the energy is 
time-dependent. In fact, an oscillatory pattern is revealed
in Sec.\ref{sec:energy-dynamics}.

\subsection{Momentum density}

\hs In an attempt to investigate the zero-momentum, as well as
higher-momentum states, we set out to compute the momentum density
of the system given by the Fourier transform

\begin{equation}
\rho_{FT}(k_x,k_y;t)\,=\,\frac{1}{4 \pi^2} 
\int_{-\infty}^{+\infty} \int_{-\infty}^{+\infty} 
\rho(x,y;t) e^{i k_x x+i k_y y} d x d y, \label{eq:FT.momentum.density}
\end{equation}

where $\rho(x,y;t)\,\equiv\,|\phi(\mathbf{r};t)|^2$ and $(k_x,k_y)$ is 
a 2D$^{\hbox{al}}$ momentum state. The goal is to explore the dynamics of 
the condensate fraction as well as the momentum density distribution.

\subsection{Numerics and initial conditions}\label{sec:numerics-and-init0}

\hs We applied a previously-written Fortran 77 code \cite{Muruganandam:2009}
to numerically solve Eq.(\ref{eq:rescaled-2DTDGPE}) in real time, using 
the split-step Crank-Nicolson method. For our purposes, we modified the
code by including the PL potential Eq.(\ref{eq:PL-potential}) and
the moving Gaussian potential 
Eq.(\ref{eq:velocity-dependent-Gaussian-potential}), instead of the 
anisotropic harmonic trap already present in the code. The simulations
were conducted on a grid of $400\times 400$ square pixels, the edge of 
each having the size of a step, 0.05. The time step was chosen to be 
$\Delta=0.001$. The code has a part which initializes the system and a
transient part during which the system evolves. In the initialization
step, the nonlinearity is introduced gradually (adiabatically) by a
stepwise introduction of the interaction parameter ${\cal N}$ at a rate
of $d{\cal N}/dt\,=\,{\cal N}/(N_{stp}\Delta)$, where $N_{stp}$ is the
number of time steps $\Delta$ during which the system is initialized.
In essence, we modified the initialization part to include a gradual
introduction of the Gaussian potential 
Eq.(\ref{eq:velocity-dependent-Gaussian-potential}) by a similar gradual
increase of its parameters. In fact, we explore in this paper two 
different initialization conditions via which the obstacle is introduced 
into the system along with the nonlinearity. It will be later shown, 
that the generation of DWs depends strongly on the initialization 
conditions of the system. We find that one condition [(i) below] leads 
to the generation of DWs, whereas the other [(ii) below] does not. 
These conditions are as follows:

\begin{itemize}
\item[i)] Simultaneously with the nonlinearity-introduction, the 
system is initially subjected to a moving obstacle whose amplitude 
$A$ and velocity $v$ are gradually increased with time from zero 
until they reach the desired specific values in this simulation.
The corresponding rates are $dA/dt\,=\,A/(N_{stp} \Delta)$ and 
$dv/dt\,=\,v/(N_{stp} \Delta)$, respectively. At the end of the 
initialization, the obstacle will have left the center of the BEC. 
Subsequently, a second equivalent obstacle is again abruptly switched 
on at the center of the BEC, with the same parameters reached at the 
end of the previous initialization. Then it is set into motion with
speed $v$ until it exits the trap.

\item[ii)] Simultaneously with the nonlinearity-introduction, stationary 
obstacle is applied in the initialization process, whose amplitude is 
gradually increased at the rate $A/(N_{stp} \Delta)$. After initialization, 
the obstacle is set into motion with speed $v$ until it exits the trap.
\end{itemize}

For the interactions and obstacle we used the parameter values 
${\cal N}\,=\,5$, $A\,=\,30$, $\beta\,=\,3$, and $v\,=\,2$. Further, 
we used $N_{stp}\,=\,42000$ such that the rates $dA/dt$ and $dv/dt$,
using $\Delta\,=\,0.001$, are $0.714$ and $0.048$, respectively. 
Similarly, ${\cal N}$ is gradually increased at a rate of 
${\cal N}/(N_{stp} \Delta)\,=\,0.119$. For $^{87}$Rb atoms with
some of the same parameters used in Ref.\cite{Ruostekoski:2001}, 
i.e., an s-wave scattering length $a_s=5.4$ nm and a trapping frequency
$\omega_{ho} = 2\pi\,\times\,25$ Hz, the trap length $\ell$ is 
$\sim 2.16 \times 10^4\,\AA$. If we choose $\lambda=100$ so that the
width of the ground state $\phi_0(z)$ becomes extremely small along 
the $z-$axis, one gets $N\sim 40$ particles. If we box length is taken
$2 L=15 \mu\hbox{m}$, then the density of our system is 
$n\sim\sim N/(4 L^2) = 4\times 10^{10}\hbox{m}^{-2}$. Consequently, 
our systems are in the dilute regime $n a_s^2 \sim 10^{-6}$. 
Ref.\cite{Ruostekoski:2001} used particle numbers of the order 
$\sim 10^3$, and thus we aim at checking the effect of using a much 
lower particle number.

\hs Experimentally, then, i) one initially sweeps a very weak laser 
beam whose intensity and velocity are gradually increasing. After the
initialization, when the initial laser beam has left the BEC, a second
laser beam is abruptly switched on at the center of the BEC. Then it
is set into motion with a constant velocity or acceleration. The second 
laser beam has the same characteristics as its predecessor at the end
of the later initialization. ii) One subjects the system to a stationary 
laser beam whose intensity is gradually increasing with time. After 
the intensity has reached a maximum, the laser beam is set into motion 
with a constant velocity or acceleration. 

\hs We therefore propose experiments in which one could try to verify
the influence of the latter conditions on the generation of DWs. 

\section{Results}\label{sec:results}

\hs In what follows, we present the results of our simulations. Our major
findings are as follows. The moving obstacle inside the BEC generates, in 
addition to vortices 
\cite{Jackson:1998,Caradoc:1999,Caradoc:2000,Gonzalez:2007,Madison:2000,Raman:2001}, 
2D$^{\hbox{al}}$ circular DWs radiating away from the moving obstacle. 
As their wave fronts reach the hard walls of the box-potential, they are 
reflected backwards upon which they interfere with incoming circular DWs. 
This interference creates interesting self-interacting matter-wave (SIMW) patterns. 
On increasing the powers $p_1$ and $p_2$ of Eq.(\ref{eq:PL-potential}) to 
values beyond 2.0, these SIMWs are suppressed. The suppression prevents the 
energy loss from being carried away by the DWs and saves the energy for 
another purpose. In the present case, this energy is directed into another 
channel in which it is expended on exciting a larger number of solitons 
inside the trap. Similarly, on increasing the anisotropy of the harmonic 
or PL trap, DWs travelling parallel to the minor axis of the trap, are 
also suppressed. We have reasons for speculating the presence of real
solitons in our systems as opposed to only solitonlike structures reported
earlier in Ref.\cite{Ruostekoski:2001}. This is largely supported by
the oscillatory behaviour of the energy dynamics indicative of soliton 
generation \cite{Parker:2010}. By inspecting the density structure at the 
center of the BEC at some time, the solitons turn out to be of a vortex 
structure. Finally, the momentum distribution of the system reveals the 
presence of a zero-momentum condensate density and the condensate fraction 
oscillates with time. We would like to draw the attention of the reader, 
that all the upcoming results have been recorded after an initialization 
process of length $t=4.2$. Thus, when for example a time $t=3$ is mentioned, 
it refers to the transient time of the simulation. The total simulation time 
is then 7.2 for this particular case. However, we do not record results in 
the initialization stage, as we would like to inspect post-initialization
phenomena. The following results were obtained using the initial condition
(i) in Sec.\ref{sec:numerics-and-init0}. It is only in 
Sec.\ref{sec:alternative-init0} that we discuss the effect of condition (ii) 
in Sec.\ref{sec:numerics-and-init0}.

\begin{figure}[t!]
\includegraphics*[width=8cm,viewport=125 414 496 709,clip]{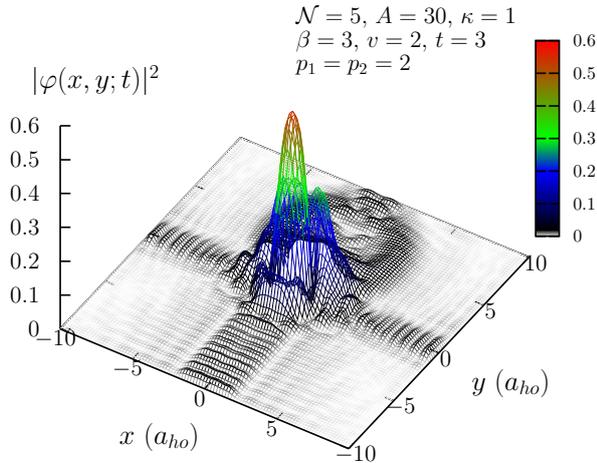}
\caption{\bls Density of a 2D$^{\hbox{al}}$ trapped BEC excited by a moving 
obstacle, Eq.(\ref{eq:velocity-dependent-Gaussian-potential}), as obtained 
by a Crank-Nicolson simulation of the 2D$^{\hbox{al}}$ time-dependent GPE, 
Eq.(\ref{eq:rescaled-2DTDGPE}), at $t=3.0$. The parameters 
used are ${\cal N}=5.0$, $p_1=p_2=2.0$, and $\kappa=1.0$. As for the obstacle, 
Eq.(\ref{eq:velocity-dependent-Gaussian-potential}), we used $A$=30, 
$\beta$=3.0, and $v$=2.0. The Fortran program used was that of Adhikari 
and Muruganandam \cite{Muruganandam:2009}. Lengths are in units of the
trap [\aho\ ].}
\label{fig:plotdensityG5A30B3V2T03Stephalf}
\end{figure}

\begin{figure}[t!]
\includegraphics*[width=8cm,viewport=101 417 497 729,clip]{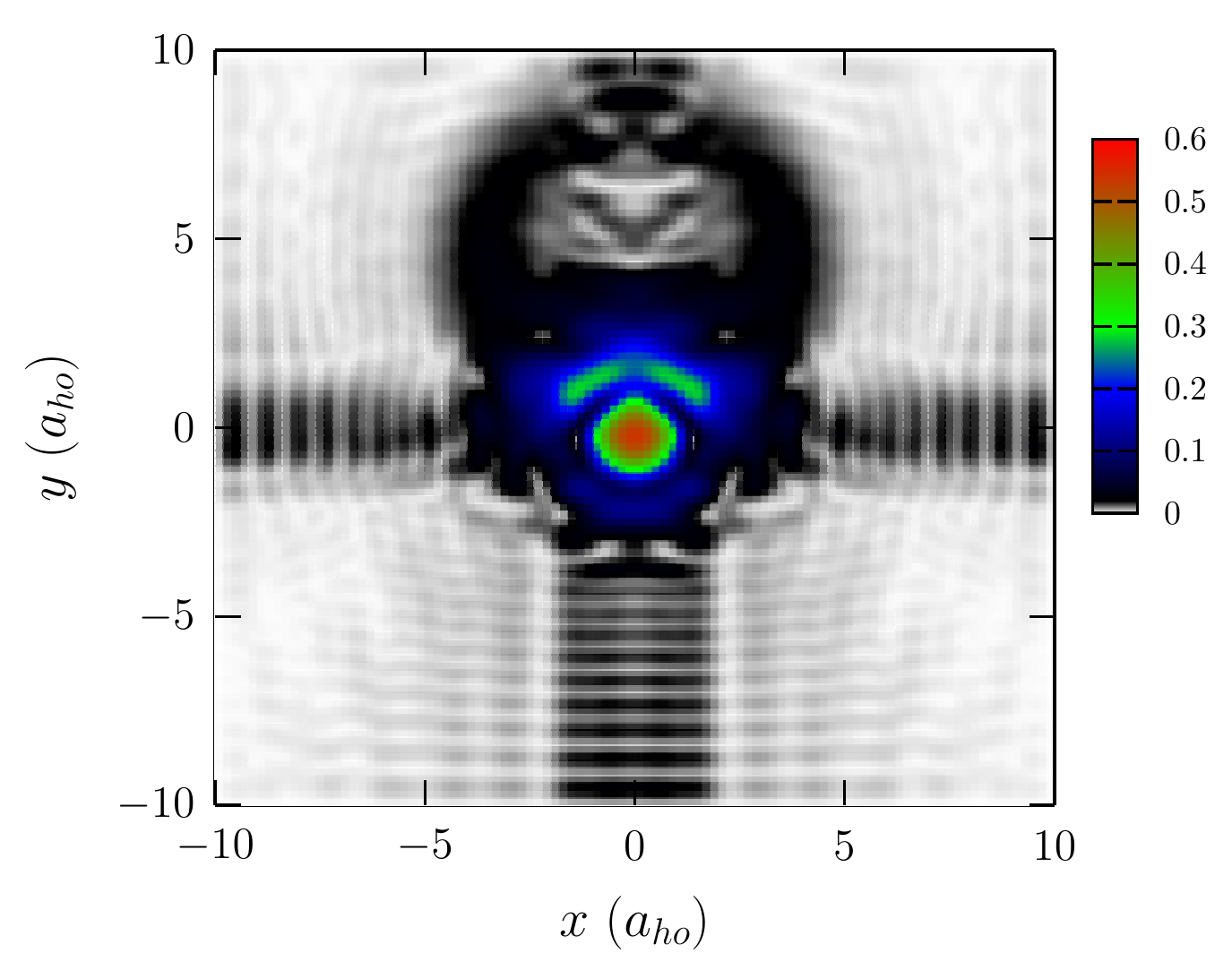}
\caption{\bls Density map of Fig.\ft\ref{fig:plotdensityG5A30B3V2T03Stephalf}. 
Lengths are in units of the trap [\aho\ ].}
\label{fig:plotdensityG5A30B3V2T03Step0d5MAP}
\end{figure}

\begin{figure}[t!]
\includegraphics*[width=8cm,viewport=127 414 496 714,clip]{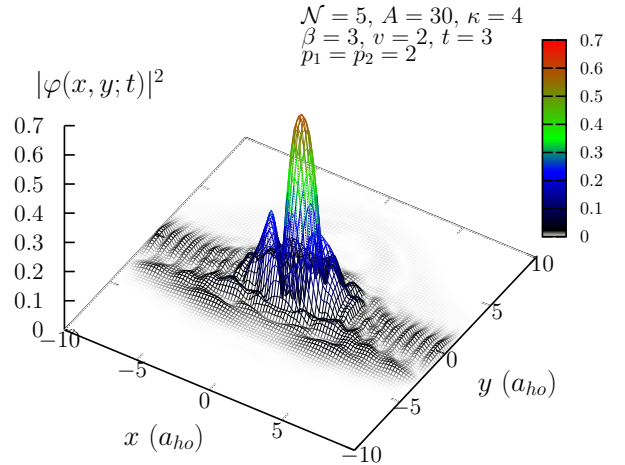}
\caption{\bls Same as in Fig.\,\ref{fig:plotdensityG5A30B3V2T03Stephalf}; but 
for an anisotropic trap with $\kappa=4.0$ [Eq.(\ref{eq:PL-potential})]. 
Lengths are in units of the trap [\aho\ ].}
\label{fig:plotdensityG5A30B3V2T03StephalfK2}
\end{figure}

\subsection{Harmonic trap with an obstacle of constant velocity}\label{sec:HOwOB}

\hs Figure\,\ft{\ref{fig:plotdensityG5A30B3V2T03Stephalf}} displays a 
2D$^{\hbox{al}}$ density plot obtained at a transient time of $t= 3$ 
for ${\cal{N}}=5.0$, $p_1=p_2=2$ (harmonic trap), and $\kappa=1.0$. 
Fig.\ft\ref{fig:plotdensityG5A30B3V2T03Step0d5MAP} displays the 
corresponding density map. For the obstacle we used $A=30$, $\beta=3.0$, 
and $v=2.0$. The intriguing result is that four, mutually perpendicular, 
2D$^{\hbox{al}}$ SIMW stripes arise. Two of them propagate parallel 
to the $x-$axis, the others parallel to the $y-$axis along the motion 
of the obstacle. In the area surrounding the latter cross-like pattern,
some residual faint interference pattern is also present. We shall relate 
this phenomenon to an artefact of the trapping geometry discussed later 
in Sec. \ref{sec:artefact-of-geometry}. It is noted that the SIMW stripes 
generated in the regime $y > 0$ are perturbed by the obstacle causing 
the density depression (small valley) in the upstream direction, behind 
the central BEC peak at $t=3$. Further, upon a closer inspection of the 
area surrounding the small moving ``valley" in the upper part of 
Fig.\ft\ref{fig:plotdensityG5A30B3V2T03Step0d5MAP}, a pair of ``holes" 
can be seen identifying a pair of vortices. This valley displays also
a complex structure which could be due to vortex excitations. It should 
be recalled that the motion of the obstacle is in the positive $y-$direction. 
For further demonstration, see the movie in the supplementary material 
section \cite{movieDD:2011} corresponding to the time evolution of 
$|\varphi(x,y;t)|^2$ in Fig.\ft\ref{fig:plotdensityG5A30B3V2T03Stephalf}.
Upon a careful inspection of the movie, one can see a solitonlike
structure in the matter-wave field upon the reflection of the BEC from
the hard walls, similar to what has been reported by 
Ref.\cite{Ruostekoski:2001}.

\hs Figure\,\ft\ref{fig:plotdensityG5A30B3V2T03StephalfK2} displays 
the 2D$^{\hbox{al}}$ density for the same parameters as in 
Fig.\ft\ref{fig:plotdensityG5A30B3V2T03Stephalf}, except that the trap
is now anisotropic with $\kappa=4$, with $x$ the major and $y$ the minor 
axis. Because of a stronger confinement along the $y-$axis, 
2D$^{\hbox{al}}$ SIMW stripes are absent in that direction, as outgoing 
matter waves from the obstacle are suppressed, unable to tunnel through 
the potential barrier. Rather, they are observed to emerge only along 
the positive $x-$axis. There is another matter wave emerging parallel 
to the 2D$^{\hbox{al}}$ SIMW in 
Fig.\ft\ref{fig:plotdensityG5A30B3V2T03StephalfK2}, which has a 
larger wavelength and therefore less energy than the accompanying SIMW. 

\begin{figure}[t!]
\includegraphics*[width=8.5cm,viewport=115 500 503 724,clip]{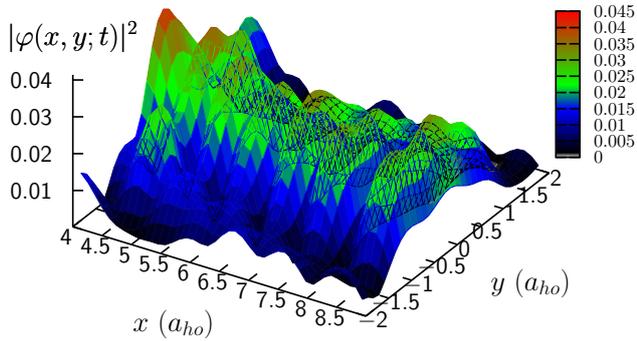}
\caption{Magnified section of the SIMW stripes appearing in 
Fig.\ft\ref{fig:plotdensityG5A30B3V2T03Stephalf}, showing how 
these waves propagate. Lengths are in trap units [\aho ].}
\label{fig:plotdensitysectiontau3G5A30B3V2powx2powy2}
\end{figure}

\begin{figure}[t!]
\includegraphics*[width=8.5cm,viewport=115 502 508 727,clip]{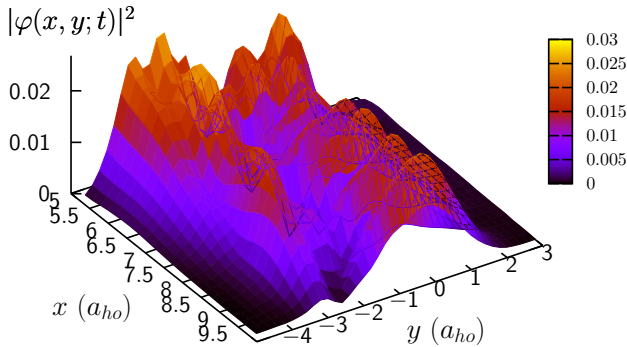}
\caption{Same as in Fig.\ft\ref{fig:plotdensitysectiontau3G5A30B3V2powx2powy2}; 
but for Fig.\ft\ref{fig:plotdensityG5A30B3V2T03StephalfK2}
with $\kappa$=4.0. There are now two separate wave structures.
One centered along the $y=0$ axis which is a SIMW; the other along the
$y=-4$ axis. Lengths are in trap units [\aho ].}
\label{fig:plotdensityG5A30B3V2T03Step5percentK2section}
\end{figure}

\hs In Fig.\ft\ref{fig:plotdensitysectiontau3G5A30B3V2powx2powy2}, a 
magnified view of a section of the SIMW stripes appearing in 
Fig.\ft\ref{fig:plotdensityG5A30B3V2T03Stephalf} is presented. 
Apparently, the SIMWs eventually propagate almost without a significant 
change in width or amplitude. Further, they display a peculiar complex 
structure. Another magnified view of a section of the SIMWs of the 
anisotropic case of Fig.\ft\ref{fig:plotdensityG5A30B3V2T03StephalfK2} 
is displayed in Fig.\ft\ref{fig:plotdensityG5A30B3V2T03Step5percentK2section}. 
This time, two separate wave structures are revealed. One is the SIMW 
centered along the $y=0$ axis; the other is a decaying matter wave of 
a larger wave length centered along the $y=-4$ axis. The latter second 
wave decays after a certain distance ($\sim$ 10 trap lengths) from 
$x=0$ and is therefore less energetic than the DWs.

\subsection{Harmonic trap with an accelerated obstacle}

\hs The purpose of this section is to demonstrate that 2D$^{\hbox{al}}$ 
SIMW stripes can still be generated by applying an accelerated obstacle 
(AO). For this purpose, a variant of the Gaussian potential 
[Eq.(\ref{eq:velocity-dependent-Gaussian-potential})] is introduced, 
where the velocity term is replaced by an acceleration term $(1/2) a t^2$:

\begin{equation}
\tilde{V}_{AO}\,=\,A \exp\left[-\beta x^2\,-\,
\beta\left(y - \frac{1}{2} a t^2\right)^2\right],
\label{eq:accelerated-moving-potential}
\end{equation}

where $a$ is the acceleration of the potential. Another goal is to 
display the effect of the ``force" of an obstacle on the dynamics 
of a 2D$^{\hbox{al}}$ trapped BEC. The obstacle has a ``mass", as 
has been outlined earlier by Astrakharchik and Pitaevskii 
\cite{Astrakharchik:2004}, and might therefore be able to produce 
a shock wave when accelerated.

\begin{figure}
\includegraphics*[width=8.5cm,viewport=127 415 496 714,clip]{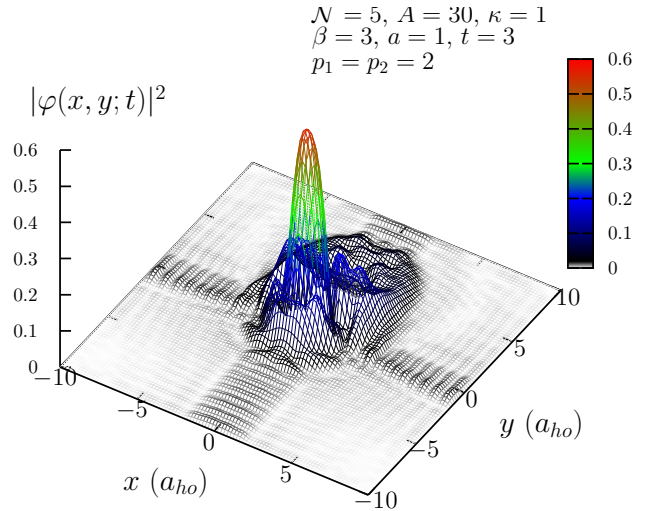}
\caption{\bls Same as in Fig.\ft\ref{fig:plotdensityG5A30B3V2T03Stephalf}; 
but with an accelerated obstacle with $a=1.0$ 
[cf. Eq.(\ref{eq:accelerated-moving-potential})].} 
\label{fig:plotdensityG5A30B3AC1T20Step5percentK1}
\end{figure}

\hs Figure \ref{fig:plotdensityG5A30B3AC1T20Step5percentK1} displays
the resulting density plot at $t = 3$ for the same parameters as in 
Fig.\ft\ref{fig:plotdensityG5A30B3V2T03Stephalf}, except that we are 
using the AO Eq.(\ref{eq:accelerated-moving-potential}) with $a=1$ instead of 
Eq.(\ref{eq:velocity-dependent-Gaussian-potential}). It can be seen that,
again similarly to Fig.\ft\ref{fig:plotdensityG5A30B3V2T03Stephalf}, 
a perpendicular set of 2D$^{\hbox{al}}$ SIMW stripes is observed. As a 
result, one concludes that an accelerated obstacle is also able to 
generate 2D$^{\hbox{al}}$ SIMW stripes in a 2D$^{\hbox{al}}$ trapped 
BEC.

\begin{figure}[t!]
\includegraphics*[width=8.5cm,viewport=121 414 502 739,clip]{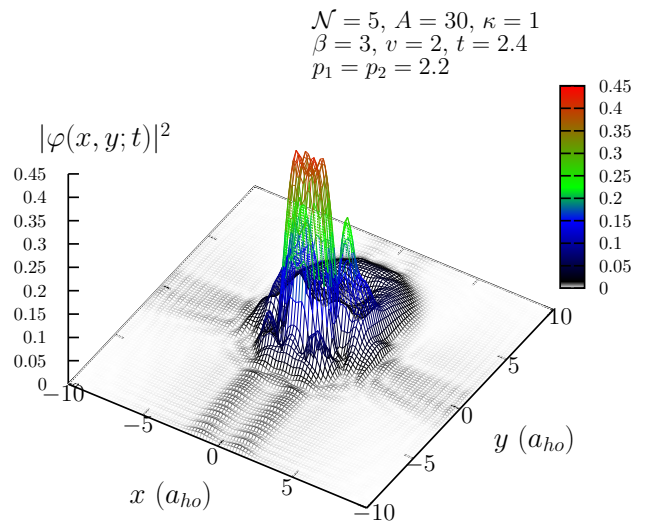}
\caption{Same as in Fig.\ft\ref{fig:plotdensityG5A30B3V2T03Stephalf}; but for 
$p_1=p_2$=2.2 at $t=2.4$. Lengths are in units of the trap [\aho ].}
\label{fig:density_plot5A30B3V2T2point4Lambda1powx2point2powy2point2step5percent}
\end{figure}

\begin{figure}
\includegraphics*[width=8.5cm,viewport=123 566 501 769,clip]{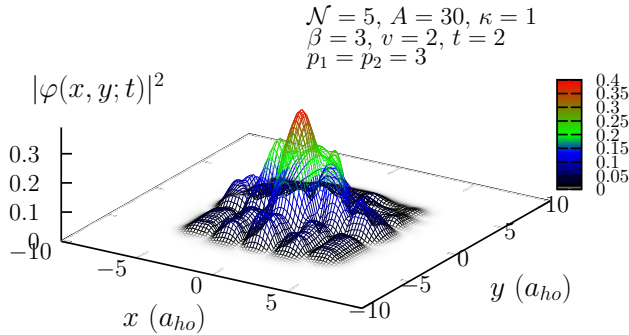}
\caption{Same as in Fig. \ref{fig:plotdensityG5A30B3V2T03Stephalf};
but for $p_1=p_2=3.0$ at $t=2.0$. Lengths are in units of the trap 
[\aho ].}
\label{fig:plotdensityG5A30B3V2T2Lambda1powx3powy3step5percent}
\end{figure}

\begin{figure}
\includegraphics*[width=8.5cm,viewport=122 417 500 734,clip]{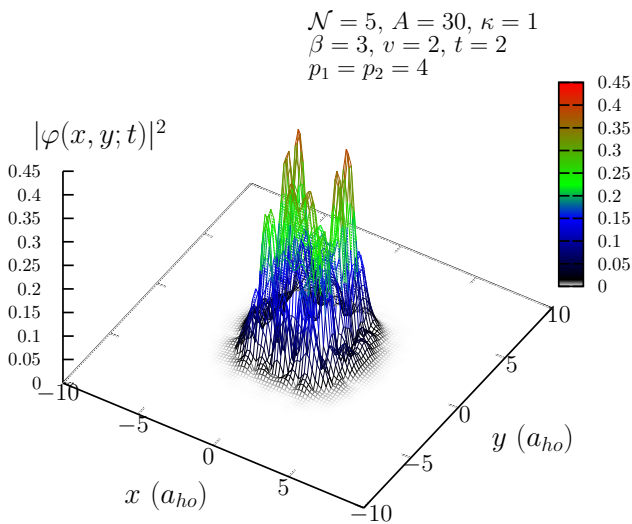}
\caption{Same as in Fig. \ref{fig:plotdensityG5A30B3V2T03Stephalf};
but for $p_1=p_2=4.0$ at $t=2.0$. Lengths are in units of the trap [\aho ].}
\label{fig:plotdensityG5A30B3V2T2Lambda1powx4powy4step5percent}
\end{figure}

\subsection{Nonharmonic potentials: effect of trap curvature}

\hs In this section, we explore whether one can still obtain 
2D$^{\hbox{al}}$ SIMW stripes if one uses a stronger PL trap
with $p_1$, $p_2\,>\,2$. It was found that for a choice of 
$p_1=p_2=2.2$, one still obtains these stripes similar to 
Fig.\ft\ref{fig:plotdensityG5A30B3V2T03Stephalf}, as displayed in 
Fig.\ft\ref{fig:density_plot5A30B3V2T2point4Lambda1powx2point2powy2point2step5percent}.
However, a closer inspection of the latter figure reveals a pair of 
{\it parallel} stripes propagating downstream along the negative 
$y-$axis. The parameters are the same as those used in 
Fig.\ft\ref{fig:plotdensityG5A30B3V2T03Stephalf}, except for $p_1$
and $p_2$. The intensity of the 2D$^{\hbox{al}}$ SIMW stripes is, 
however, smaller parallel to the $x-$axis than the $y-$axis. On 
going up to larger powers, $p_1=p_2\ge 3$, the 2D$^{\hbox{al}}$ SIMW
stripes are suppressed as displayed in 
Figs.\ft\ref{fig:plotdensityG5A30B3V2T2Lambda1powx3powy3step5percent}
and \ref{fig:plotdensityG5A30B3V2T2Lambda1powx4powy4step5percent}.
The curvature at the edges of the trap plays a crucial role in the 
suppression of the SIMWs as explained later in Sec. \ref{sec:discussion}.

\begin{figure}[t!]
\includegraphics*[width=8.5cm,viewport=191 524 556 771,clip]{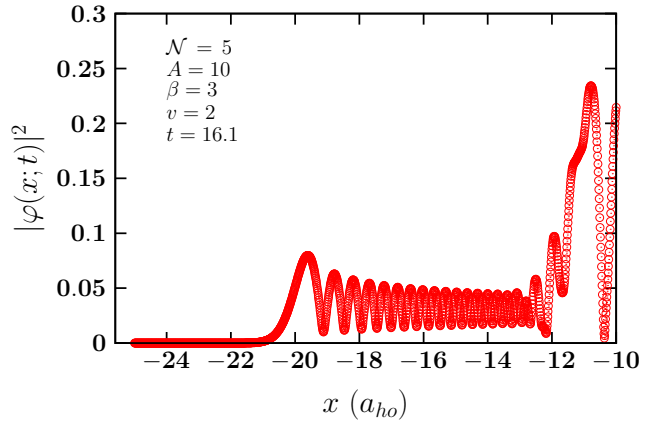}
\caption{One-dimensional SIMWs generated by a 1D$^{\hbox{al}}$
Crank-Nicolson simulation of the time-dependent GPE 
Eq.(\ref{eq:1D-version-of-the-TDGPE}). The system is a 1D$^{\hbox{al}}$ 
hard-sphere Bose gas of ${\cal N}=5$. The vortex potential parameters 
are $A=10$, $\beta=3$, and $v=2$. Parameters and lengths are in 
trap units [\aho ].}
\label{fig:plotdensity1D.G5A10B3V2}
\end{figure}

\subsection{One-dimensional simulations}

\hs In this section, we present additional results of the Crank-Nicolson
simulation of the 1D$^{\hbox{al}}$ GPE, i.e., the 1D$^{\hbox{al}}$ version 
of Eq.(\ref{eq:rescaled-2DTDGPE}):

\begin{equation}
\left[-\frac{1}{2}\frac{\partial^2}{\partial x^2}\,+\,\tilde{V}(x;t)\,
+\,{\cal N}|\varphi(x;t)|^2\,-\,i\frac{\partial}{\partial t}\right]
\varphi(x;t)\,=\,0,
\label{eq:1D-version-of-the-TDGPE}
\end{equation}

where 

\begin{equation}
\tilde{V}(x;t)\,=\,\frac{1}{2} x^2\,+\,A \exp[-\beta(x\,-\,vt)^2],
\end{equation}

and the velocity $v$ is directed along the $x-$axis. Further, as in the
2D$^{\hbox{al}}$ case, the harmonic trap is cut off by a 1D HWBP such that

\begin{equation}
\varphi(-L_x;t)\,=\,\varphi(L_x;t)\,=\,0.
\end{equation}

The goal is to provide additional support for our arguments made about 
the nature of the observed 2D$^{\hbox{al}}$ SIMW stripes. 

\hs Figure\ft\ft\ref{fig:plotdensity1D.G5A10B3V2} demonstrates the 
density $|\varphi(x;t)|^2$ at a time $t = 16.1$ for 
${\cal N}=5$, $A=10$, $\beta=3$, and $v=2$. One can see an SIMW 
emitted to the left of the central BEC density travelling along 
the $-x$ direction. The sharp dip in the density near $x=-10$ is
a dark soliton excited by the moving obstacle. The DW was able to 
climb up the external potential barrier up to a distance of $\sim 20$ 
trap lengths from the center of the trap! Amazingly, these waves are 
largely energized, enough to travel that far from the center of the 
trap. Given for example a distance of $x\,=\,20 a_{ho}$, one estimates 
this wave must have an energy of at least $200 \hbar\omega_{ho}$. 
Hence, a question arises whether the moving obstacle is indeed the 
mechanism which would impart such a high energy to these excitations. 
In fact, if one were to compute the ``mass" of the obstacle and 
calculate the associated kinetic energy, one would then verify this 
fact. Thus, let us make the following estimation: If the ``mass" of 
an obstacle is $m$ and its velocity $v$, then by assuming the obstacle 
imparts all its kinetic energy to the outgoing 1D$^{\hbox{al}}$ 
dispersive waves from the trap center, these waves need an energy of 
$(1/2) mv^2\sim 200\hbar\omega_{ho}$ to climb up the external harmonic 
barrier a distance $x=20a_{ho}$, before getting reflected back from the 
box potential. Accordingly, if $v=2$, then $m\sim 100 m_B$, where 
$m_B$ is the boson mass incorporated in the trap length 
$a_{ho}=\sqrt{\hbar/(m_B \omega_{ho}})$. That is, the obstacle posesses
a large mass equivalent to the mass of $\sim 100$ bosons, all moving 
at $v=2$. Given the latter mass, one can conclude that the moving 
obstacle is able to impart very high energies to the system, exciting 
dispersive waves able to surmount the external potential barrier to 
extremely large distances!

\begin{figure}[t!]
\includegraphics*[width=8.5cm,viewport=170 524 499 774,clip]{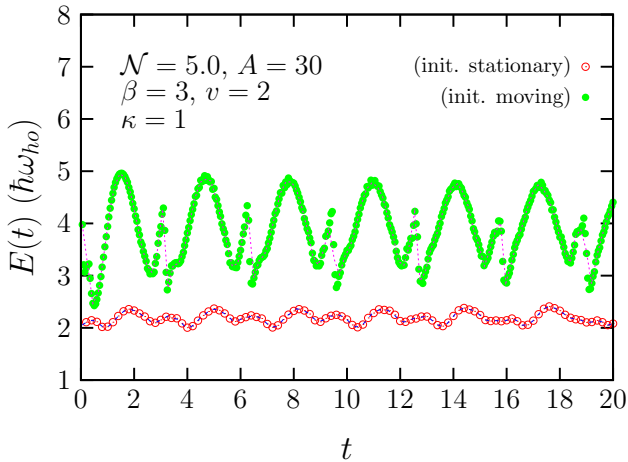}
\caption{Dynamics of the energy $\langle E(t)\rangle$ 
[Eq.(\ref{eq:energy-functional})] of the system in 
Fig.\ft\ref{fig:plotdensityG5A30B3V2T03Stephalf} [solid circles, 
initial condition (i)]. The open circles are for 
Fig.\ft\ref{fig:plotdensityG5A30B3V2powx2powy2init0} below 
with initial condition (ii). Energies are in trap units 
($\hbar\omega_{ho}$).}
\label{fig:plotEvstG5A30B3V2powx2powy2init0}
\end{figure}

\subsection{Energy Dynamics}\label{sec:energy-dynamics}

\begin{figure}[t!]
\includegraphics*[width=8.5cm,viewport=122 503 501 758,clip]{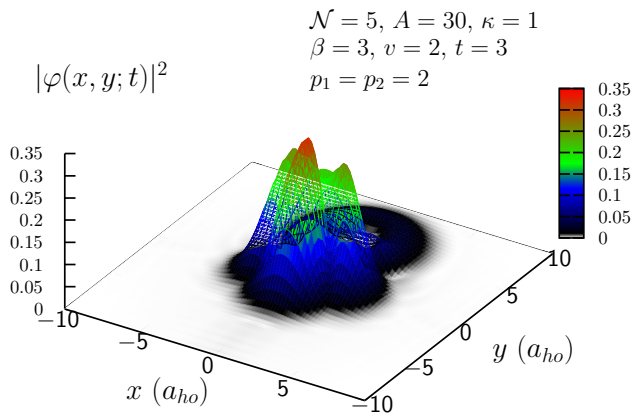}
\caption{As in Fig.\ft\ref{fig:plotdensityG5A30B3V2T03Stephalf}; but
using the initial conditions (ii) in Sec.\ref{sec:numerics-and-init0}.}
\label{fig:plotdensityG5A30B3V2powx2powy2init0}
\end{figure}

\hs In this section, we display the time-dependence of the energy
functional [Eq.(\ref{eq:energy-functional})] for the systems 
considered in the previous sections. The goal is to present 
evidence for a possible existence of solitons inside the 
2D$^{\hbox{al}}$ trapped BEC excited by a moving obstacle,
based on the fact that the time-dependence of the energy is 
oscillatory for solitons \cite{Parker:2010}. First, we present
Fig.\ft\ref{fig:plotEvstG5A30B3V2powx2powy2init0} which displays the energy 
dynamics for the system of Fig.\ft\ref{fig:plotdensityG5A30B3V2T03Stephalf} 
[solid circles, initial condition (i)]. The solid circles are for
Fig.\ft\ref{fig:plotdensityG5A30B3V2powx2powy2init0} below with 
initial condition (ii) added here for the purpose of comparison.
The energy $E(t)$ [Eq.(\ref{eq:energy-functional})] fluctuates 
in time about a seemingly stable time average; this is indicative 
of the presence of solitons in the system \cite{Parker:2010}. 
It can also be argued that the system is emitting and reabsorbing
energy. 

\begin{figure}[t!]
\includegraphics*[width=8.5cm,viewport=163 521 499 773,clip]{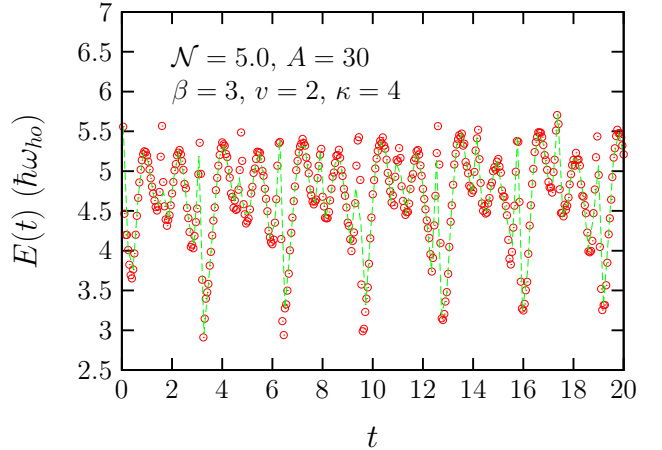}
\caption{As in Fig.\ft\ref{fig:plotEvstG5A30B3V2powx2powy2init0};
but for the system in 
Fig.\ft\ref{fig:plotdensityG5A30B3V2T03StephalfK2}. Energies are in 
units of the trap $\hbar\omega_{ho}$.}
\label{fig:plotEvstG5A30B3V2T20Step5percentK2}
\end{figure}

\begin{figure}[t!]
\includegraphics*[width=8.5cm,viewport=163 521 499 773,clip]{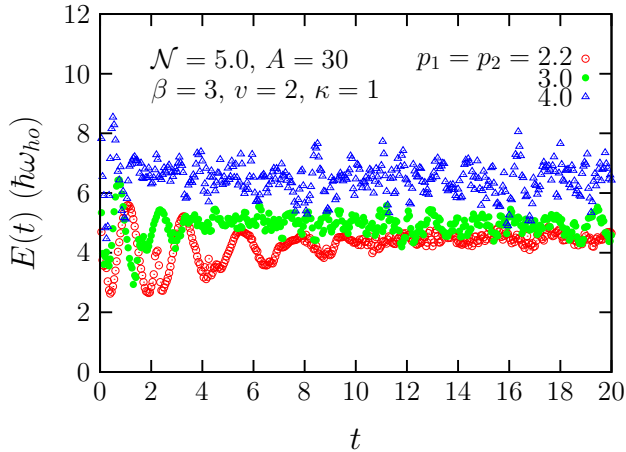}
\caption{Dynamics of the energy $\langle E(t)\rangle$ 
[Eq.(\ref{eq:energy-functional})] for various trapping
geometries: open circles: $p_1=p_2=2.2$, system of 
Fig.\ft\ref{fig:density_plot5A30B3V2T2point4Lambda1powx2point2powy2point2step5percent};
solid circles: $p_1=p_2=3$, system of 
Fig.\ft\ref{fig:plotdensityG5A30B3V2T2Lambda1powx3powy3step5percent};
open triangles: $p_1=p_2=4.0$, system of 
Fig.\ft\ref{fig:plotdensityG5A30B3V2T2Lambda1powx4powy4step5percent}.
Energies are in units of the trap $\hbar\omega_{ho}$.}
\label{fig:plotEvstG5A30B3V2T20Step0d05variousPows}
\end{figure}

\hs Figure\ft\ref{fig:plotEvstG5A30B3V2T20Step5percentK2} displays
the energy dynamics for the anisotropic system of 
Fig.\ft\ref{fig:plotdensityG5A30B3V2T03StephalfK2}, and
Fig.\ft\ref{fig:plotEvstG5A30B3V2T20Step0d05variousPows} the energy
dynamics for the systems in the different trapping geometries of 
Figs.\ft\ref{fig:density_plot5A30B3V2T2point4Lambda1powx2point2powy2point2step5percent}-\ref{fig:plotdensityG5A30B3V2T2Lambda1powx4powy4step5percent}, with
$p_1=p_2=2.2$ (open circles), $p_1\,=\,p_2\,=3.0$ (solid circles),
and $p_1=p_2=4$ (open triangles), respectively. 
Fig.\ft\ref{fig:plotEvstG5A30B3V2T20Step5percentK2} reveals a more rapid
and spikier behavior in the energy oscillations than the other two
figures. The system in an anisotropic trap thus tends to emit and 
absorb energy at a faster rate than in an isotropic trap. 
Fig.\ft\ref{fig:plotEvstG5A30B3V2T20Step0d05variousPows} displays
the effect of the trap curvature on the energy oscillatory patterns.
Note that the oscillatory behavior becomes more irregular as the
power of the trap is increased via $p_1$ and $p_2$. In fact, the 
energy seems to become almost stable with time when $t>4$ for
$p_1=p_2=2.2$ and 3.

\hs Next we proceeded to determine the role of the moving obstacle 
and the nonlinearity of the TDGPE in the observed energy oscillations. 
We therefore checked the time-dependence of the energy, 
Eq.(\ref{eq:energy-functional}) for two cases: 1) in the absence of 
the obstacle and the presence of the nonlinearity and 2) vice versa. 
In case (1) the system reached equilibrium immediately after the 
initialization process and stopped evolving. No SIMWs were observed
whatsoever as the BEC is neither excited nor is the magnetic trap
switched off to allow its expansion. The energy and density pattern 
stabilized with time as well. No energy oscillations were observed. 
However, in case (2), SIMWs were generated. The system and energy 
were still evolving after the initialization process and the energy 
displayed oscillations. Therefore, the main source for the observed 
oscillations is the moving obstacle. Consequently, the moving obstacle 
not only generates the 2D$^{\hbox{al}}$ SIMW stripes observed in 
Figs.\ft\ref{fig:plotdensityG5A30B3V2T03Stephalf},
\ref{fig:plotdensityG5A30B3V2T03StephalfK2},
\ref{fig:plotdensityG5A30B3AC1T20Step5percentK1}, and
\ref{fig:density_plot5A30B3V2T2point4Lambda1powx2point2powy2point2step5percent}-\ref{fig:plotdensityG5A30B3V2T2Lambda1powx4powy4step5percent},
but also solitons inside the central part of the BEC. 

\begin{figure}[t!]
\includegraphics*[width=8.5cm,viewport=111 501 515 765]{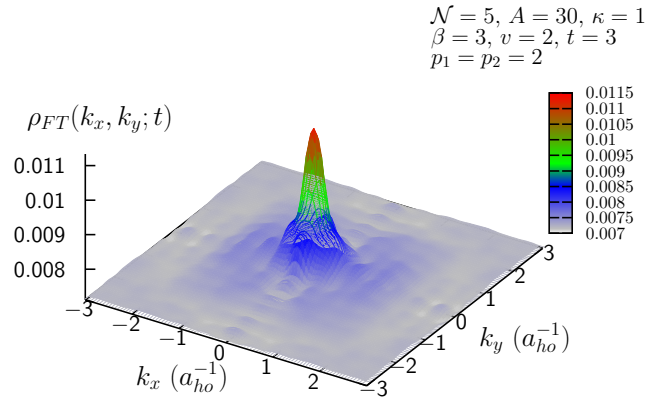}
\caption{Momentum density distribution of the system in
Fig.\ft\ref{fig:plotdensityG5A30B3V2T03Stephalf} obtained
by the Fourier transform Eq.(\ref{eq:FT.momentum.density}).
Wave vectors $k_x$ and $k_y$ are in units of the trap 
($a_{ho}^{-1}$).}
\label{fig:plotdensityFTtau3G5A30B3V2powx2powy2}
\end{figure}

\begin{figure}[t!]
\includegraphics*[width=8.5cm,viewport=111 500 518 763]{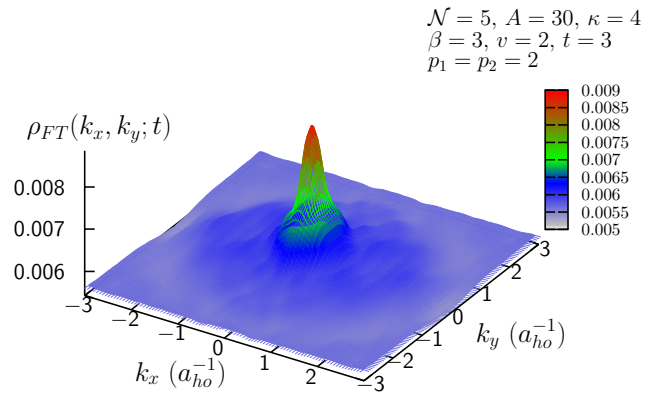}
\caption{As in Fig.\ft\ref{fig:plotdensityG5A30B3V2T03Stephalf}; but
for the system of Fig.\ft\ref{fig:plotdensityG5A30B3V2T03StephalfK2}.}
\label{fig:plotdensityFTtau3G5A30B3V2Kappa4powx2powy2}
\end{figure}

\subsection{Alternative initial conditions}\label{sec:alternative-init0}

\hs Here, we demonstrate the effect of using the alternative 
initial condition (ii) presented in Sec. \ref{sec:numerics-and-init0}. 
This is when the laser potential (obstacle) is initially kept 
stationary, while its intensity is gradually increased up to 
a certain chosen maximum. After this maximum is reached, it is set into
motion. The resulting density at $t=3$ is shown in 
Fig.\ft\ref{fig:plotdensityG5A30B3V2powx2powy2init0} which is 
practically the same as Fig.\ft\ref{fig:plotdensityG5A30B3V2T03Stephalf}, 
but using the alternative initial condition (ii) in 
Sec.\ref{sec:numerics-and-init0}. The previously-observed 
SIMW pattern is not obtained at $t = 3$, even at later times 
(not shown). This result holds similary for the anisotropic case 
tackled earlier in Fig.\ft\ref{fig:plotdensityG5A30B3V2T03StephalfK2}. 
Consequently, one can argue a very peculiar result, which shows 
that generation of SIMWs is dependent on the initial conditions of 
an excited BEC. We would also like to note that the energy of this
system displays an oscillatory behaviour as well (as shown in
Fig.\ft\ref{fig:plotEvstG5A30B3V2powx2powy2init0} (open circles)). 
Henceforth, the observed energy oscillations are not a result of 
specific initialization conditions.

\subsection{Dynamics of momentum density}

\hs Here we try to detect the condensate in order to explore its
dynamics. For this purpose, the time dependent momentum density is 
computed from the Fourier transform of the spatial density 
$\rho(x,y;t)$, as given by Eq.(\ref{eq:FT.momentum.density}).

\begin{figure}[t!]
\includegraphics*[width=8.5cm,viewport=111 500 518 767]{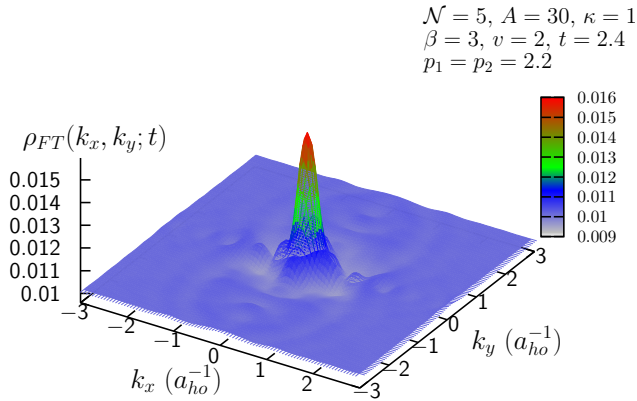}
\caption{As in Fig.\ft\ref{fig:plotdensityG5A30B3V2T03Stephalf};
but for the system of 
Fig.\ft\ref{fig:density_plot5A30B3V2T2point4Lambda1powx2point2powy2point2step5percent}.}
\label{fig:plotdensityFTtau3G5A30B3V2powx2d4powy2d4}
\end{figure}

\hs Figures \ref{fig:plotdensityFTtau3G5A30B3V2powx2powy2}-\ref{fig:plotdensityFTtau3G5A30B3V2powx2d4powy2d4} 
display the momentum density $\rho_{FT}(k_x,k_y;t)$ of 
Figs.\ft\ref{fig:plotdensityG5A30B3V2T03Stephalf},
\ref{fig:plotdensityG5A30B3V2T03StephalfK2}, and
\ref{fig:density_plot5A30B3V2T2point4Lambda1powx2point2powy2point2step5percent},
respectively, and at the indicated times $t$. In all the latter 
momentum density figures, there exists a Gaussian-like, 
zero-momentum density peak centered at $(k_x,k_y)\equiv (0,0)$. 
The smoothness of this peak and the surrounding density in $k-$space 
indicate that the BEC excitations in 
Figs.\ft\ref{fig:plotdensityFTtau3G5A30B3V2powx2powy2}-\ref{fig:plotdensityFTtau3G5A30B3V2powx2d4powy2d4} occur in a
continuous band of $(k_x,k_y)$ states. In fact the regime around
the central BEC peak, so to speak, displays a nonzero momentum
density surface. By inspecting 
Fig.\ft\ref{fig:plotdensityFTtau3G5A30B3V2powx2powy2} closely,
one can observe some rectangular-like momentum-density wave 
structure surrounding the central BEC peak. In other words, the
effects of the geometry-artefact (see Sec. \ref{sec:artefact-of-geometry}) 
are manifested even in the $k-$space of the current system.
Fig.\ft\ref{fig:plotdensityFTtau3G5A30B3V2Kappa4powx2powy2}
shows no specific structure around the central peak, but rather
a smooth momentum density distribution. Nevertheless, and upon
a close inspection of the momentum density along the $k_x=\pm3$
lines in the neighborhood of $k_y=0$, some depression in the
density is observed as indicated by a slight bending in the plane.
Fig.\ft\ref{fig:plotdensityFTtau3G5A30B3V2powx2d4powy2d4} on the
other hand displays some valleyed structure in the vicinity of the 
zero-momentum peak.

\begin{figure}[t!]
\includegraphics*[width=8.5cm,viewport=216 500 500 772]{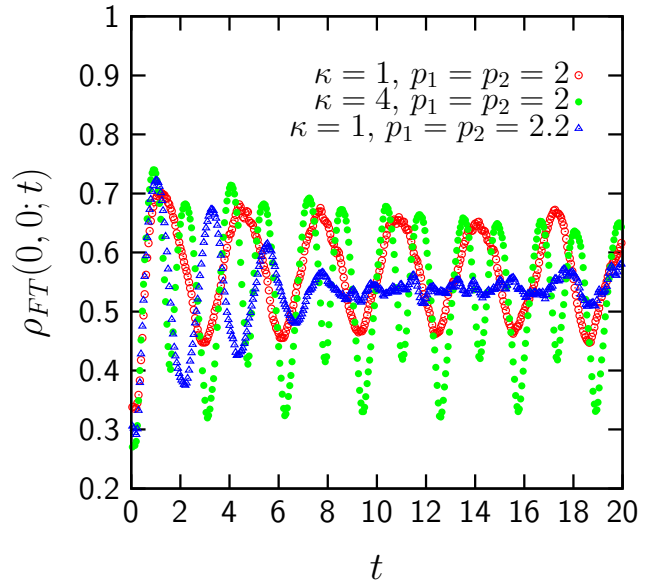}
\caption{Dynamics of the zero-momentum density $\rho(0,0;\tau)$
for the systems of Fig.\ft\ref{fig:plotdensityG5A30B3V2T03Stephalf} 
with $\kappa=1$, $p_1=p_2=2$ (open circles); 
Fig.\ft\ref{fig:plotdensityG5A30B3V2T03StephalfK2} with $\kappa=4$,
$p_1=p_2=2$ (solid circles); and 
Fig.\ft\ref{fig:density_plot5A30B3V2T2point4Lambda1powx2point2powy2point2step5percent} 
with $\kappa=4$, $p_1=p_2=2.2$ (open triangles),
respectively.}
\label{fig:plot.condensate.dynamics.several.traps}
\end{figure}

\hs Throughout the evolution of the system, the central BEC
peak neither disappears nor does it expand significantly, 
although it fluctuates in amplitude. Consequently, the zero-momentum 
condensate density $\rho_{FT}(0,0;t)$ displays oscillatory patterns 
shown in 
Fig.\ft\ref{fig:plot.condensate.dynamics.several.traps} 
corresponding to the latter systems.
Fig.\ft\ref{fig:plotdensityFTtau3G5A30B3V2powx2powy2}: (open circles)
$\kappa=1$, $p_1=p_2=2$; 
Fig.\ft\ref{fig:plotdensityFTtau3G5A30B3V2Kappa4powx2powy2}:
(solid circles) $\kappa=4$, $p_1=p_2=2$; 
Fig.\ft\ref{fig:plotdensityFTtau3G5A30B3V2powx2d4powy2d4}:
(open triangles) $\kappa=1$, $p_1=p_2=2.2$.
The oscillations in $\rho_{FT}(0,0;t)$ are a strong indication 
to particles being excited out of the condensate as 
$\rho_{FT}(0,0;t)$ decreases, and particles ``falling back" 
into the condensate as $\rho_{FT}(0,0;t)$ rises again. 
Therefore the presence of SIMW phenomena seems to promote 
excitations out of and deexcitations back into the condensate. 
For further visual demonstration of the momentum dynamics of 
the system, see the movie in the supplementary material section 
\cite{movieFT:2011}, which corresponds to 
Fig.\ft\ref{fig:plotdensityFTtau3G5A30B3V2powx2powy2}. Further,
one must note that the condensate oscillations are substantially 
damped in the range $8 \stackrel{<}{\sim} t \stackrel{<}{\sim} 18$ 
for $p_1 = p_2 = 2.2$.

\section{Discussion}\label{sec:discussion}

\hs We now set out to find an explanation for the cross-like 
2D$^{\hbox{al}}$ SIMW stripes observed in
Figs.\ft\ref{fig:plotdensityG5A30B3V2T03Stephalf}, 
\ref{fig:plotdensityG5A30B3AC1T20Step5percentK1}, and
\ref{fig:density_plot5A30B3V2T2point4Lambda1powx2point2powy2point2step5percent}.
In addition, we discuss briefly the work of Ruostekoski \ea\ 
\cite{Ruostekoski:2001}, the analogy with 1D$^{al}$ dispersive
waves, backflow, and a possible existence for vortices and solitons.

\subsection{Work of Ruostekoski \ea\ \cite{Ruostekoski:2001}}

\hs Ruostekoski \ea\ \cite{Ruostekoski:2001} explored the 
self-interference of BEC matter-waves in hard-wall traps of several 
geometries. Their work is very much related to ours and it provides
significant results. According to these authors, the hard walls can 
be realized by the application of blue-detuned far-off resonant
laser-light sheets. Their goal was to study the evolution of a BEC upon 
the reflection of these matter-waves from the hard-walls. Particularly, 
the nonlinear Talbot effect was investigated in a 1D hard-wall trap, 
where solitonlike structures were obtained. It was noted, that the
nonlinear Talbot effect of a coherent matter-wave field is analogous
to the Fresnel diffraction effect in optics. Similarly, we can 
argue that the cross-like matter-wave pattern observed in our
Fig.\ft\ref{fig:plotdensityG5A30B3V2T03Stephalf} above, is also
analogous to the Fraunhofer diffraction pattern \cite{vanderPoel:2001}.

\hs In general, they initially confined the system in a magnetic field 
trap added to the hard-wall potential. Upon switching off the magnetic 
trap, the BEC expanded and was reflected from the hard-wall boundaries 
causing a self-interference pattern. Ruostekoski \ea\ conducted these 
investigations in (i) a 1D box with an isotropic magnetic trap; (ii) a 
3D$^{\hbox{al}}$ highly anisotropic cigar trap with hard-walls introduced 
by laser-light sheets in the axial direction of the trap; (iii) a 
2D$^{\hbox{al}}$ circular boundary, and (iv) a 2D$^{\hbox{al}}$ square 
box, both of them with isotropic traps. 
In (i), a complex self-interference pattern is obtained in the time
evolution of the BEC density, characterized by canals corresponding to 
the evolution of solitonlike structures. This is where the Talbot effect 
was observed. The ``density holes" observed in the density-dynamics 
corresponded to a gray solitonlike structure. A similar 
solitonlike structure is obtained in the second case (ii) of the cigar
trap, though with some differences. In case (iii), the system is investigated
for two subcases: initially symmetric, and with an initial broken symmetry.
In the former ring solitons are observed, in the latter the solitonlike
structures break and form vorticity. In (iv), the reflections of the
matter-waves result also in solitonlike structures.

\hs The main difference between our work and that of Ruostekoski
\ea\ is that we do not switch off the magnetic trap, but rather we
excite the BEC by a moving laser-potential. In addition, we needed
to apply certain initial conditions in order to cause an excitation
of dispersive waves. Another difference is, whereas the latter authors
observe the Fresnel diffraction effect in the 2D$^{\hbox{al}}$ square box potential,
we are able to obtain cross-like matter-wave patterns analogous to
a Fraunhofer diffraction pattern. In addition, we would like to
emphasize that we have focused here on the energy and momentum
dynamics of the BEC matter-wave field.

\subsection{Artefact of the Geometry}\label{sec:artefact-of-geometry}

\hs According to Ref.\cite{Ruostekoski:2001}, then, the rectangular pattern 
of the waves observed in Figs.\ft\ref{fig:plotdensityG5A30B3V2T03Stephalf},
\ref{fig:plotdensityG5A30B3V2T03StephalfK2},
\ref{fig:plotdensityG5A30B3AC1T20Step5percentK1}, and 
\ref{fig:density_plot5A30B3V2T2point4Lambda1powx2point2powy2point2step5percent}
is due to the nature of the boundary conditions imposed on the wave
function at the edges of the square mesh during the simulation.
Our boundary conditions Eq.(\ref{eq:boundary-conditions}) force the 
wave function to vanish there, thus mimicing an impenetrable hard
wall. According to Eq.(\ref{eq:total-trapping-geometry}), one can see 
that into this box there is embedded the 2D$^{\hbox{al}}$ 
harmonic oscillator (or PL) trap, both of which share the same 
center. The hard walls of the box are parallel to the coordinate 
axes $x$ and $y$ and the harmonic (or PL) trap {\it ends} where 
the hard walls begin and cannot reach beyond. Yet, the motion of 
the obstacle energizes the system to such an extent producing waves 
capable of reaching the edges of the square mesh, but not farther 
than this. At the hard walls the wave function of the system 
vanishes and $-$possessing enough energy$-$ it is reflected from 
the HWBP. As a result, a self-interfering matter-wave pattern forms, 
as waves outgoing from the trap center interfere with ``incoming" 
waves reflected from the hard walls. Inspect also carefully our 
movie in the supplementary section \cite{movieDD:2011}. The 
perpendicular 2D$^{\hbox{al}}$ SIMWs observed are a result of a trapping 
geometry-artefact arising from the inner curvature of the magnetic 
trap (PL trap), the HWBP, and the finite width of the central BEC 
peak from which the outgoing DWs are emitted. The presence of a moving 
density dip due to the obtacle is largely responsible for the presence 
of this cross-like SIMW pattern. Further, one can argue similarly to 
Ref.\cite{Ruostekoski:2001} that the latter cross-like patterns is 
a result of an initial symmetry breaking of the system caused by an 
initial slight displacement via the moving obstacle. On the other hand, 
the pair of parallel stripes observed in 
Fig.\ft\ref{fig:density_plot5A30B3V2T2point4Lambda1powx2point2powy2point2step5percent} 
is most likely due to the emission of a pair of nonconcentric circular DWs 
from the central BEC peak, each of which is reflected from the hard wall, 
creating its own SIMW pattern.

\hs Now when the trap is strengthened by increasing its power
exponents, i.e., by setting $p_1$ and $p_2$ larger than 2.2 in
Eq.(\ref{eq:PL-potential}), the expanding BEC waves do not 
reach the HWBP anymore and rather decay by tunneling through 
the edges of the PL trap. As a result, no SIMW patterns are formed in 
Figs.\ft\ref{fig:plotdensityG5A30B3V2T2Lambda1powx3powy3step5percent} 
and \ref{fig:plotdensityG5A30B3V2T2Lambda1powx4powy4step5percent}.

\subsection{Analogy with 1D$^{\hbox{al}}$ dispersive waves}

\hs We consider an investigation by Hakim \cite{Hakim:1997}, which 
carries some similar features to ours. He explored the flow of a 
nonlinear Schr\"odinger fluid past an immobile obstacle in a one 
dimensional homogeneous Bose gas and found that the obstacle 
repeatedly emitted gray solitons propagating downstream, i.e., 
opposite to the motion of the fluid. At the same time he observed 
upstream propagating dispersive waves, i.e., in the direction of 
fluid flow. We can, therefore, explain the initial emission of waves 
from the center of the trap, and before they are reflected from the 
HWBP, based on an analogy with Hakim's investigation. Hakim 
\cite{Hakim:1997} demonstrated the excitation of dispersive waves 
(DWs) in a 1D$^{\hbox{al}}$ uniform Bose gas in the direction of obstacle motion 
or fluid flow. Basing on Hakim's results, we could then argue that 
the waves emitted from the trap center in our 2D$^{\hbox{al}}$ system are similarly 
circular DWs. Consequently, an outgoing circular DW interferes with an 
``incoming" linear DW reflected off the box-potential boundaries. However, 
the cross-like pattern observed is a peculiar result which, to the best 
of our knowledge, has not been reported elsewhere.

\subsection{Vortices and solitons}

\hs On inspecting Fig.\ft\ref{fig:plotdensityG5A30B3V2T03Stephalf} closely,
one can see two cylindrical, mantel-like walls surrounding the central BEC
peak. The region between each ``mantel" and the central BEC peak displays
a deep density depression which is reminiscent of crescent vortex 
solitons \cite{He:2008}. There is a mantel in the front of the central 
peak (in the region $y<0$) and in the back (at $y>0$). In addition, the 
pair of ``holes" which was indicated to earlier in Sec.\ref{sec:HOwOB},
is a vortex anti-vortex pair similar to the one observed by Jackson \ea\
\cite{Jackson:1998} except that in our pair the vortices are much smaller.
It is anticipated that vortex excitations could play a role in exciting DWs
in 2D$^{\hbox{al}}$ systems. Indeed, the oscillating feature of the energy dynamics shown in 
Figs.\ft\ref{fig:plotEvstG5A30B3V2powx2powy2init0}-\ref{fig:plotEvstG5A30B3V2T20Step0d05variousPows}
is a manifestation of solitonic presence inside the trap. According to 
Parker \ea\ \cite{Parker:2010}, the energy of a soliton is oscillatory; 
this supports our previous conclusion regarding the possible presence of 
vortex solitons at some instant of time in 
Fig.\ft\ref{fig:plotdensityG5A30B3V2T03Stephalf}.

\subsection{Suppression of dispersive-waves and soliton generation}

\hs Dispersive waves are suppressed in an anisotropic trap along
the strongly-confining direction, as displayed in 
Fig.\ft\ref{fig:plotdensityG5A30B3V2T03StephalfK2}; whereas they
are still emitted along the major axis of the elongated trap. The
dispersive waves do not have enough kinetic energy to climb up
the external potential barrier along the strongly-confining direction
(the minor axis of the trap).

\hs Next to this, upon increasing the curvature of the isotropic 
confinement, as in 
Figs.\ft\ref{fig:density_plot5A30B3V2T2point4Lambda1powx2point2powy2point2step5percent}-\ref{fig:plotdensityG5A30B3V2T2Lambda1powx4powy4step5percent},
the number of solitons inside the trap increases significantly. The
suppression of DWs prevented the energy necessary for solitonic 
excitations to be carried away by dispersion and be expended
on climbing up the external potential barrier. One can conclude 
that a larger number of solitons can be generated in a 
strongly-confining PL trap by a moving obstacle.

\subsection{Backflow}

\hs According to Feynman and Cohen (1953) (FC) \cite{Glyde:1994}, the 
motion of an obstacle in a fluid causes backflow of the atoms
in the neighborhood of the moving obstacle. Referring back to the
FC theory, an excited atom $i$ in a strongly interacting Bose fluid
such as liquid \xhe4\ is surrounded by adjacent atoms $j$ that correlate
with atom $i$ and are thus able to flow around it. The correlation
function can be found in Ref.\cite{Glyde:1994}. As such, one could
further note that collective excitations in our current system generate 
waves travelling away from the center of the trap which are later  
reflected from the 2D$^{\hbox{al}}$ box-potential. We further believe
that in the initial condition (i) in Sec.\ref{sec:numerics-and-init0}, 
the gradually introduced laser potential induces a backflow pattern 
which, at the end of the initialization stage, interacts with the backflow 
pattern of the secondary laser potential introduced in the transient stage
of the simulation. This is in a manner consistent with the backflow physics
discussed in an earlier investigation by Ghassib and Chatterjee 
\cite{Ghassib:1983}, where they computed the interaction potential
between the backflow patters resulting from the motion of two impurities
in a non-viscous fluid in various geometries.

\subsection{Future work}

\hs The numerical analysis of the current system conducted by 
applying the Crank Nicolson method to the 2D$^{\hbox{al}}$ TDGPE 
lead us to specific information on the dynamics of this system 
as elaborated on in Sec. \ref{sec:results}. However, other information
regarding the quantum hydrodynamics of this system is of importance
and needs to be investigated. Particularly the phase of the wavefunction 
could reveal further information about the coherence properties. 
For this purpose, the Madelung transformation (MT) 
\cite{Nonnenmacher:1983} can be utilized in obtaining two analytic 
equations arising from the real and complex part of the MT. In this 
regard, the dynamic phase of the wavefunction reveals information
about the dynamic superfluid velocity field in the presence of the 
moving obstacle potential. In the future, we will investigate also
trapped BECs excited by an attractive (red-detuned) laser potenial.

\section{Conclusion}\label{sec:conclusion}

\hs In summary, then, we have demonstrated the creation of
highly energetic, two-dimensional, self interfering matter-wave (SIMW) 
stripes inside a 2D$^{\hbox{al}}$ BEC confined by a power-law (PL) potential trap 
cut off by hard-wall box-potential (HWPB) boundaries. The split-step
Crank Nicolson method was used to numerically solve the time-dependent
Gross-Pitaevskii equation (TDGPE) using a code previously developed
by Muruganandam and Adhikari \cite{Muruganandam:2009}. It was found
that the SIMWs are generated by the interference of circular dispersive 
waves (DWs) emitted from the center of the trap and their reflected 
wave fronts off the box-potential walls. Four mutually perpendicular 
SIMW stripes are observed, two of them parallel to the obstacle's motion 
along the $y-$axis and the others perpendicular to $y$. As we have 
indicated, the DWs are chiefly excited by an obstacle moving inside 
the 2D$^{\hbox{al}}$ BEC under specific conditions. These conditions 
include an interaction strength of the order of ${\cal N}\sim 1$ and 
harmonic confinement, or a PL trap, given by Eq.(\ref{eq:PL-potential}) 
with $p_1$, $p_2\stackrel{<}{\sim} 2.2$.

\hs The suppression of dispersive waves by strong confinement results in
a buildup of energy inside the BEC, which eventually leads to a larger
number of soliton excitations than when this energy is expended on
DW excitations. By Fourier transforming the spacial density distribution, 
the momentum density dynamics displayed a nonvanishing, oscillating 
zero-momentum condensate density. As the strength of confinement was 
increased, the latter oscillations began to disappear in a certain
time range, as revealed in Fig.\ft\ref{fig:plot.condensate.dynamics.several.traps}
for $p_1=p_2=2.2$, indicating that the condensate fraction began to stabilize
somewhat in that regime. Particularly, the condensate density oscillations
signal the excitation and deexcitation of particles from and to the
zero-momentum state. Further, a vortex anti-vortex pair was obtained
in the vicinity of the moving laser obstacle similar to, but smaller
in size, to those of Jackson \ea\ \cite{Jackson:1998}.

\hs In essence, then, we presented in this regard a 
phenomenological examination of these systems in analogy to earlier 
examinations by Hakim \cite{Hakim:1997} and He \ea\ \cite{He:2008}. 
The excited waves were classified as dispersive based on the earlier 
1D$^{\hbox{al}}$ investigation of Hakim \cite{Hakim:1997}, who found 
that a nonlinear Schr\"odinger fluid flowing past a stationary obstacle 
can emit dispersive waves. By considering the inverse of this argument, 
i.e., a moving obstacle inside a stationary BEC, we arrived at the latter 
classification. In addition, following somewhat Hakim's investigation, we 
conducted 1D$^{\hbox{al}}$ simulations for a BEC trapped in a harmonic potential 
and were able to observe density oscillations excited in the BEC. These 
oscillations shown in Fig.\ft\ref{fig:plotdensity1D.G5A10B3V2} are 
dispersive waves.  

\acknowledgements
\hs The authors are grateful to The University of Jordan for
partially supporting this research under project number
74/2008-2009 dated 19/8/2009. One of us (H.B.G.) is grateful
to The University of Jordan for granting him a sabbatical
leave in the academic year 2010/2011 during which the present
work was undertaken under the general title ``Many-Body
Systems: Further Studies and Calculations [Part Two]".

\bibliography{vortex.bib}
\end{document}